\documentclass[prd,a4paper,aps,twocolumn,showpacs,
               nofootinbib,nobibnotes,superscriptaddress]
              {revtex4}

\usepackage{epsfig}
\usepackage{amsmath}
\usepackage{amssymb}

\newcommand{\ie}{{\em i.e. }}
\newcommand{\eg}{{\em e.g. }}

\newcommand{\ra}{\rightarrow}

\newcommand{\gsim}{\gtrsim}
\newcommand{\lsim}{\lesssim}
\newcommand{\dd}{\partial}

\newcommand{\De}{\Delta}

\newcommand{\om}{\omega}

\newcommand{\be}{\begin{equation}}
\newcommand{\ee}{\end{equation}}
\newcommand{\bee}{\begin{equation*}}
\newcommand{\eee}{\end{equation*}}
\newcommand{\bea}{\begin{eqnarray}}
\newcommand{\eea}{\end{eqnarray}}
\newcommand{\bean}{\begin{eqnarray*}}
\newcommand{\eean}{\end{eqnarray*}}
\newcommand{\ba}{\begin{array}}
\newcommand{\ea}{\end{array}}

\newcommand{\bk}{{\mathbf k}}

\newcommand{\bx}{{\mathbf x}}

\newcommand{\ti}{t_{\rm in}}
\newcommand{\tf}{t_{\rm fin}}

\newcommand{\ds}{\displaystyle}

\begin{document}

\title{On the frequency of gravitational waves}

\author{Chiara Caprini}
\email{chiara.caprini@physics.unige.ch}
\affiliation{D\'epartement de Physique Th\'eorique, Universit\'e de
  Gen\`eve, 24 quai Ernest Ansermet, CH--1211 Gen\`eve 4, Switzerland}

\author{Ruth Durrer}
\email{ruth.durrer@physics.unige.ch}
\affiliation{D\'epartement de Physique Th\'eorique, Universit\'e de
  Gen\`eve, 24 quai Ernest Ansermet, CH--1211 Gen\`eve 4, Switzerland}

\author{Riccardo Sturani}
\email{riccardo.sturani@physics.unige.ch}
\affiliation{D\'epartement de Physique Th\'eorique, Universit\'e de
  Gen\`eve, 24 quai Ernest Ansermet, CH--1211 Gen\`eve 4, Switzerland}
\affiliation{INFN, Italy}
\date{\today}

\begin{abstract}
We show that there are physically relevant situations where
gravitational waves do not inherit the frequency
spectrum of their source but its wavenumber spectrum.
\end{abstract}


\maketitle

\section{Introduction}
Let us consider a source of (weak) gravitational waves (GWs), an anisotropic 
stress $\Pi_{ij}(\bx,t)$. The induced GWs
can be calculated via the linearized Einstein equations. In the
case of an isolated, non-relativistic, far away source, one can derive the
quadrupole formula (see e.g.~\cite{straumann}), $r=|\bx|, c=1$
\be
h_{ij}(\bx,t) = \frac{2G}{r}\ddot Q_{ij}(t-r)~,
\ee
where $Q_{ij}$ denotes the quadrupole of the source, and we are considering
the perturbed metric
$g_{\mu\nu}=\eta_{\mu\nu}+h_{\mu\nu}$. The wave has the
same time dependence as the source. If we
have an isolated source which has a harmonic time dependence with
frequency $\omega_s$, $\Pi_{ij}(\bx,t) = \Pi_{ij}(\bx)e^{-i\om_s t}$, the wave 
zone approximation  gives, far away from the source,
\be
h_{ij}(\bx,t) = \frac{4Ge^{-i\om_s (t-r)}}{r}
    \int d^3x'\,\Pi_{ij}(\bx')e^{-i\om_s \hat{\bx}\cdot\bx'}~.
\label{wavezone} \ee Again, $h_{ij}$ has inherited the frequency of
the source: it is a spherical wave  whose amplitude in direction
$\hat\bx$ is determined by $\tilde\Pi_{ij}(\bk)$ with 
$\bk=\om_s\hat\bx$. As we will show in this brief report, this
simple and well known fact from  linearized general relativity and,
equivalently, electrodynamics, has led to  some errors 
when applied to GW sources  of
cosmological origin. 

As an example we consider a first order phase transition in the
early universe. This can lead to a period of turbulent motion in the
broken phase fluid, giving rise to a GW signal which
is in principle observable by the planned space interferometers
LISA or BBO
\cite{kamionkowski,kosowsky,dolgov,notari,dolgovgrasso,our,sargent}. 
One can describe this phase of turbulence in the fluid as
a superposition of turbulent eddies: eddies of characteristic size
$\ell$ rotate with frequency $\om_\ell \simeq v/\ell\neq 1/\ell$, since 
$v<1$. The wavenumber spectrum of the eddies peaks at $k_L\simeq 1/L$, where
$L$ is the stirring scale, and has the usual Kolmogorov shape for
$k\gsim 1/L$ ($\ell<L$). The above considerations about GW
generation prompted several workers in the 
field~\cite{kamionkowski,kosowsky,dolgov,notari,dolgovgrasso}
to infer that the induced GWs will directly inherit the
frequency of the eddies. Especially, even if the turbulent spectrum in
$k$-space peaks at wave number $k_L\simeq 1/L$, 
Refs. \cite{kamionkowski,kosowsky,dolgov,notari,dolgovgrasso} conclude that 
the GW
energy density spectrum peaks instead at the largest eddy frequency $k=\omega_L
\neq k_L$. The argument to reach this conclusion is simple and seems, at
first sight, convincing: one may consider each eddy as a 
source in the wave zone which oscillates at a given frequency. The
GWs it produces have therefore the same frequency. Adding the
signal from all the eddies incoherently, one obtains a GW
energy density spectrum which peaks at the same frequency as the frequency
spectrum of the eddies, namely at $\om=\om_L$. Since GWs obey the
dispersion relation $k=|\om|$, the $k$-space distribution of the
GWs then cannot agree with the $k$-space distribution
of the source. In this brief report we show why this simple argument is not 
correct. 
In the above case, the GW energy spectrum actually peaks at
$\om=k_L$ and the $k$-space distribution (but not the frequency
distribution) of the source and the GWs agree \cite{our}.

Thus, the result (\ref{wavezone}) has to be 
used with care in the case of a cosmological GW source which is active only
for a
finite amount of time and is spread over the entire  
universe. In Section \ref{stocha} we discuss several cases of cosmological
sources to which this consideration applies. This is relevant
especially in view of future searches for a stochastic GW
background which might be able to detect some of the features predicted   
for GW spectra of primordial origin (in particular, in the case of 
turbulence after a phase transition, the peak frequency)~\cite{sargent}.

The rest of the paper is organized as follows. In Section~II we
analyze an academic example of a source which can be represented as a
Gaussian wave packet both, in $k$-space and in frequency space. We show
when the wave will inherit the wave number of the source
and when its frequency. We then discuss the case of a generic source. In 
Section~III we consider  
a typical cosmological source: stochastic, statistically homogeneous and
isotropic, and short lived. There we also explain the subtleties which
are responsible for the above mentioned misconception in the case of
turbulence, and we discuss other examples of GW sources for which our findings
are relevant. The final section contains our conclusions.

\section{The Gaussian packet}

Let us consider a scalar field $h(\bx,t)$ fulfilling the 
usual wave equation with source $\Pi$ (for the sake of simplicity, 
we omit the tensorial structure which is irrelevant for the problem we
want to address here): 
\be\label{heq}
\Box h(\bx,t)=-16\pi G\,\Pi(\bx,t)\,.
\ee 
Here $\Box = -\dd_t^2+\De$. In terms of the Fourier transform of the
source,
\be
\tilde\Pi(\bk,\om) = \int dt \int\, d^3x\,e^{i(\bk\cdot\bx-\om t)}\Pi(\bx,t)\,,
\ee
the retarded solution of Eq.~(\ref{heq}) is ($k\equiv|\bk|$)
\be
\label{hsol}
h(\bx,t)=\frac{iG}{\pi^2}\int d^3k\frac{e^{-i\bk\cdot\bx}}k
\left[\tilde\Pi(\bk,-k)e^{-ikt}-\tilde\Pi(\bk,k)e^{ikt}\right]\,.
\ee

To illustrate the possible behavior of this solution, we now consider a
source whose Fourier transform has a Gaussian  profile both in wavenumber 
and frequency,~\footnote{The source (\ref{gaupi}) is not real in
  physical space;  to make it real one should symmetrize it in
  $\om_s$. But we can as well consider a complex source whose
  real part represents the physical source.}
\be 
\label{gaupi}
\tilde\Pi(\bk,\omega)=\Pi_0\exp\left[-\frac{(k-k_s)^2}{2\Delta_k^2}\right]
\exp\left[-\frac{(\omega-\omega_s)^2}{2\Delta_\omega^2}\right]\,,
\ee
where $k_s$ and $\om_s$ represent the characteristic frequency and
wavenumber of this source ($k_s$ is positive, while
$\om_s$ can be either positive or negative). Inserting (\ref{gaupi}) in
solution (\ref{hsol}) one immediately obtains 
\be \label{doublegau}  
\ba{c}\ds
h(\bx,t)=i\frac{4G}{\pi}\frac{\Pi_0}{r}\int_0^{\infty}dk\sin(kr)
\exp\left[-\frac{(k-k_s)^2}{2\Delta_k^2}\right]\times\vspace{5pt}\\
\!\!\!\!\ds
\left[e^{-ikt}\exp\left(-\frac{(k+\omega_s)^2}{2\Delta_\omega^2}\right) 
-e^{ikt}\exp\left(-\frac{(k-\omega_s)^2}{2\Delta_\omega^2}\right)\right]\,,
\ea
\ee
where $r\equiv|\bx|$.
The resulting wave is a superposition of plane waves with \emph{frequency}
$\pm k$. To estimate which is the dominant frequency of this superposition, 
\ie the frequency at which the amplitude peaks, one has to look at the
maximum of the double Gaussian in (\ref{doublegau}). In particular for a
source which has $\Delta_\omega\ll\Delta_k$, we expect the spectrum to be 
peaked at $k\simeq\omega_s$: we recover result (\ref{wavezone}), namely a 
spherical wave that oscillates with the characteristic frequency of
the source. In the opposite case $\Delta_\omega\gg\Delta_k$,
the spectrum is peaked at $k\simeq k_s$, and the wave inherits the
wavenumber of the source.  

This simple academic example provides us insight about the behavior of
more realistic 
sources. Note that the spacetime distribution of the above wave-packet
is again a Gaussian packet with spatial extension $\De_x = 1/\De_k$
and temporal spread $\De_t = 1/\De_\om$. A typical
astrophysical source of GWs is
long lived and has a very distinct frequency (supernovae, binary
systems). On the other hand it is well 
localized in space, so we can expect the ordinary 
$\Delta_\omega\ll\Delta_k$ case. Note however, that the signal is
suppressed if $k_s$ is very far from $\pm\om_s$. For a typical binary
we have $\om_s \simeq vk_s$, where $v$ is the velocity of the system
and it is well known that fast binaries emit a stronger GW
signal than slow ones.

On the other hand, consider a cosmological source which is
spread over all of space in a statistically homogeneous and isotropic way, to
respect the symmetries of the Friedmann universe.
This source is very extended but it may be active only over a
short period of time. In a Friedmann universe, sources can
be switched on and off everywhere at approximately the same time,
\eg at a given temperature.
This kind of sources may then fall in the $\Delta_\omega\gg\Delta_k$
range. In the extremal case of a source which is 
a delta function in time, $\De_\om \ra
\infty$, the frequency spectrum of the generated radiation is
entirely determined by the structure of the source in $k$--space.

The goal of the next section is to present an example of such a source. 
As a preparation, let us first consider a generic source
$\Pi(\bx,t)$ which is active from some initial time $\ti$ to a final
time $\tf$. Using the retarded Green's function in $k$--space
\be
G_R(k,t)=4\pi\Theta(t)\frac{\sin(kt)}{k}\,,
\label{retgreen}
\ee
one finds the solution to Eq.~(\ref{heq}) at time $t>\tf$
\be
h(\bk,t)=A\frac{\sin(k(t-\tf))}k+B\cos(k(t-\tf))\,,
\label{free}
\ee
with
\be
\ba{rcl}
\ds A(\bk)&=&\ds16\pi G\int_{\ti}^{\tf} dt'\cos(k(\tf-t'))\Pi(\bk,t')
\vspace{5pt}\\
\ds B(\bk)&=&
\ds16\pi G\int_{\ti}^{\tf} dt' \frac{\sin(k(\tf-t'))}{k}\Pi(\bk,t')\,.
\ea
\ee
If the source has some typical frequency $\om_s$ satisfying 
$(\tf-\ti)\om_s\gg 1$, the
integrals $A$ and $B$ approximate delta functions of $k-\om_s$ and
$k+\om_s$. However, frequencies of the order of $(\tf-\ti)^{-1}$ or
smaller are not especially amplified and only a detailed calculation can
tell where the resulting GW spectrum peaks.

\section{A stochastic, homogeneous and isotropic, short lived source}
\label{stocha}

\begin{figure}

\centering
\epsfig{figure=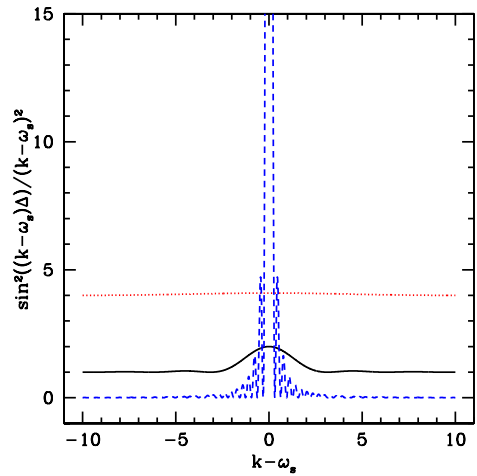,width=6.5cm} 
\caption{The quantity $\sin^2((k-\om_s)\De)/(k-\om_s)^2$, $\De=(\tf-\ti)/2$, is
  shown as a 
  function of $k-\om_s$ for the values of $\De=$ 1 (black, solid) 0.3
  (red dotted) and 10 (blue, dashed). For clarity the red (dotted) and
  black (solid) curves are displaced.\label{fig:spec}}
\end{figure}

We now analyze the case of a typical GW source of 
primordial origin: a stochastic
field, statistically homogeneous and isotropic. 
Let us consider the two point function of the spatial Fourier transform of 
the anisotropic stress tensor
\be
\langle\Pi_{ij}(\bk,t)\Pi_{kl}^*(\bk',t') \rangle=
\delta(\bk-\bk')M_{ijkl}(k,t,t')\,,\label{sourcecorr}
\ee
and analogously for the induced GW
\be
\langle h_{ij}(\bk,t) h_{kl}^*(\bk',t) \rangle=
\delta(\bk-\bk')H_{ijkl}(k,t)\,,
\ee
where we write $H_{ijkl}(k,t)$ for $H_{ijkl}(k,t,t)$. 
For simplicity we consider the wave equation (\ref{heq}) in flat 
space-time, and ignore the tensor structure of $H$ and $M$. 
From the above definitions and from Eq.~(\ref{heq}), one finds that  
in Fourier space the relation between the spectral power of the energy 
momentum tensor and that of the GW is 
\be
H(k,\omega,\omega')=(16\pi G)^2
\frac{M(k,\omega,\omega')}{(\omega^2-k^2)({\omega'}^2-{k'}^2)}\,.
\ee
Fourier transforming over $\om$ and $\om'$ and taking into account
the correct boundary and initial conditions $H\ra 0$ for $t\ra
-\infty$, one obtains
\be\label{stoch:gen}
H(k,t)=\frac{(8\pi G)^2}{k^2}\mbox{Re}
\left[M(k,k,k)-e^{-2ikt}M(k,k,-k)\right]\,,
\ee
where we have used that $M(k,\om,\om') =M^*(k,-\om,-\om')$. 
This result can also be obtained via convolution of $M(k,t,t')$ with
the Green's functions $G_R(k,t)G_R(k,t')$ of Eq.~(\ref{retgreen}). To interpret
Eq.~(\ref{stoch:gen}) let us consider a source which is active from 
some initial time $\ti$ to a final time $\tf$ and is oscillating with
frequency $\om_s$,
\be\label{stoch:ex}
M(k,t,t') = \left\{\begin{array}{ll}
 F(k)e^{i\om_s(t-t')} &\mbox{ if }\quad t\in [\ti,\tf] \\
 0  &\mbox{ if }\quad t \not\in [\ti,\tf]\,.
\end{array}\right.
\ee
Inserting the Fourier transform of this somewhat simplistic source in the above
Eq.~(\ref{stoch:gen}) one finds
\be\label{stoch:res}
H(k,t)=(16\pi G)^2\frac{F(k)}{k^2}
\frac{\sin^2[(k-\om_s)(\tf-\ti)/2]}{(k-\om_s)^2}\,,
\ee
plus terms which average to zero over an oscillation period.
The square root of the last factor in this expression becomes 
$\pi\delta(k-\omega_s)$ when $\omega_s(\tf-\ti) \ra \infty$. Hence, if 
$\om_s(\tf-\ti)\gg 1$, the resulting GW spectrum $H(k,t)$ is peaked at
$\om_s$, the typical 
frequency of the source. However, if $\om_s(\tf-\ti)\sim\mathcal{O}(1)$, the 
last term does not significantly influence the spectrum, which can then be 
peaked there where $F(k)$ is.    
The typical frequency of the wave is then 
given by the typical wave number of the source.
Note however, that at very high wavenumbers, $(\tf-\ti)(k-\om_s)\gg 1$
the last factor in Eq.~(\ref{stoch:res}) always leads to suppression. 
This behavior is illustrated in Fig.~\ref{fig:spec}.

Even though this flat space result cannot be taken over literally in
a cosmological context, since we neglected the expansion of the universe,
let us discuss in its lights, at least qualitatively, some examples of
cosmological GW sources. 

If {\bf turbulence} is generated in the early universe during a first order 
phase transition, as discussed in the introduction, 
one has the formation of a cascade of eddies. The largest eddies have a period 
comparable to the time duration of the turbulence itself (of the phase 
transition). They are not spatially correlated, and are described by a
wavenumber power spectrum which is peaked at a scale corresponding to
their size.   
According to Eq.~(\ref{stoch:res}), these eddies generate GWs which
inherit their wavenumber spectrum, since the last factor in this
equation is flat enough, not to significantly influence the
spectrum. Smaller eddies instead have higher frequencies,
corresponding to periods smaller than the duration of the
turbulence. According to Eq.~(\ref{stoch:res}), smaller eddies  can
imprint their frequency on the GW spectrum. However, they only
contribute to the slope of the GW spectrum at high $k$: the peak of
the GW spectrum is always dominated by the largest eddies. Therefore,
the GW spectrum is peaked at a wavenumber corresponding to the size of
the largest eddies. 

{\bf Primordial magnetic fields} are another cosmological source of
GWs. They simply evolve by flux conservation and dissipation of energy at
small scales, and have no characteristic oscillation frequency. 
Their lifetime is typically from their
creation in the early universe until matter domination (after which only 
a negligible amount of GWs is produced~\cite{caprini,pedro}). 
Also in this case the wavenumber is imprinted in the induced GW
spectrum. The suppression for $k (\tf-\ti)\gg 1$ is seen by the fact that
GWs are generated mainly on super-horizon scales, $kt\lsim 1$. Once a scale 
enters the Hubble horizon, further GW generation is
negligible and the produced wave evolves simply by redshifting its
frequency.

A similar case is represented by GWs produced by the 
{\bf neutrino anisotropic stresses}, which generate a turbulent 
phase~\cite{dolgov2} or simply evolve by neutrino free streaming 
~\cite{bashinsky}. In the first case, the source is short
lived and therefore its typical wavenumber is imprinted on the GW
spectrum. In the second case there is no typical frequency and, like
for magnetic fields, the relevant time is the time until a given
wavenumber enters the horizon, $kt\simeq 1$. Again, the
GW frequency corresponds to the source's wave number.

A counter example is the stochastic GW background produced by 
{\bf binaries of primordial black holes} which are copiously produced in some
models with extra-dimensions~\cite{bhb}. In this case, one has a
well defined typical frequency and the sources are long lived, so that
$\om_s(\tf-\ti)\gg 1$. The GWs inherit the frequency of the source.

\section{Conclusions}

\begin{figure}

\centering
\epsfig{figure=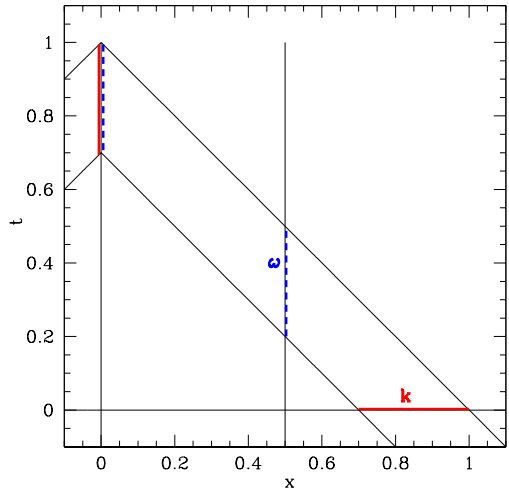,width=7cm} 
\caption{ The past light cone (on which gravitons propagate) of a
  detector at $x=0$ observing for a given time interval 
  sweeps a portion of space in the temporally confined source (solid,
  red line).  Therefore, the wave number of the source is registered
  as the frequency of the emitted GW observed at $x$. On the other
  hand, the detector observes the time evolution of a spatially
  confined source (dashed, blue line), and therefore sees its
  frequency. 
  \label{lc}}
\end{figure}

To elucidate further our findings, let us consider again a
spatially homogeneous source with well defined wave number $k$,
which is alive only for a
brief instant in time (a typical cosmological source). 
To determine the frequency of the generated
GW, we observe it for some period of time (which is much longer than the 
inverse of the frequency). 
Since the source is active only for a very short period, as time
goes on we observe GWs generated in different positions: 
what we see as happening in time, actually corresponds to a
graviton-snapshot of the source in space, taken at the instant of time when it
was on (see the bold solid lines in Fig.~\ref{lc}).

Conversely, for a source which occupies a very small spatial extension but 
lives for a long time (\eg a binary star system), the situation is inverted. 
The GW signal arriving at the observer comes
from virtually always the same position, but from different times. In
this case the GW really monitors the time evolution of
the source and therefore acquires its frequency  (see the bold dashed lines
in Fig.~\ref{lc}). 

This duality between $k$ and $\om$ is actually very natural, since they
are related by Lorentz boosts. In general, both the spatial and time structure
of a source lead to the generation of GWs.

In this brief report we have shown that the standard lore that GWs
oscillate with the same frequency as their source is not
always correct. Special care is needed in the cosmological context, where the
universe has the same temperature everywhere and can undergo some
drastic change simultaneously (but in a uncorrelated way) in a brief interval
of time. This can lead to sources of GWs which are
very extended in space, but short lived. Moreover, GWs from sources
which only have a very slow time dependence and have no typical
frequency, inherit usually the spectrum of the source in $k$-space.

In general, a source can have a complicated
spectrum with structure in both $k$-space and frequency space. If the
situation is not as clear cut as in the Gaussian wave packet, a
detailed calculation is needed to obtain the correct GWs spectrum. 
On the top of that, as one sees explicitely in Eq.~(\ref{hsol}), only the
values $(\bk,\om)$ with $|\bk|=|\om|$ of the Fourier transform of
the source contribute to the GW amplitude.

\section*{Acknowledgment}
We thank Florian Dubath, Stefano Foffa,  Arthur Kosowsky, Michele Maggiore, Gonzalo
de Palma and Geraldine Servant for stimulating 
discussions. This work is supported by the Swiss National Science
Foundation.

\end{document}